\documentclass[conference]{IEEEtran}
\IEEEoverridecommandlockouts

\usepackage{cite}
\usepackage{amsmath,amssymb,amsfonts}
\usepackage{algorithmic}
\usepackage{graphicx}
\usepackage{textcomp}
\usepackage{xcolor}
\usepackage{siunitx}
\usepackage{multirow}
\def\BibTeX{{\rm B\kern-.05em{\sc i\kern-.025em b}\kern-.08em
    T\kern-.1667em\lower.7ex\hbox{E}\kern-.125emX}}
\begin{document}

\title{A Mixed-Signal Photonic SRAM-based High-Speed Energy-Efficient Photonic Tensor Core with Novel Electro-Optic ADC \vspace{-0.0 in} \\
}

\author{\IEEEauthorblockN{Md Abdullah-Al Kaiser$^*$, Sugeet Sunder$^\dagger$, Ajey P. Jacob$^\dagger$, and Akhilesh R. Jaiswal$^*$}
\IEEEauthorblockA{\textit{$^*$University of Wisconsin--Madison, WI, USA, $^\dagger$USC Information Sciences Institute, CA, USA} \\
mkaiser8@wisc.edu, sunder@isi.edu, ajey@isi.edu, akhilesh.jaiswal@wisc.edu}
}

\maketitle

\begin{abstract}
The rapid surge in data generated by Internet of Things (IoT), artificial intelligence (AI), and machine learning (ML) applications demands ultra-fast, scalable, and energy-efficient hardware, as traditional von Neumann architectures face significant latency and power challenges due to data transfer bottlenecks between memory and processing units. Furthermore, conventional electrical memory technologies are increasingly constrained by rising bitline and wordline capacitance, as well as the resistance of compact and long interconnects, as technology scales. In contrast, photonics-based in-memory computing systems offer substantial speed and energy improvements over traditional transistor-based systems, owing to their ultra-fast operating frequencies, low crosstalk, and high data bandwidth.
Hence, we present a novel differential photonic SRAM (pSRAM) bitcell-augmented scalable mixed-signal multi-bit photonic tensor core, enabling high-speed, energy-efficient matrix multiplication operations using fabrication-friendly integrated photonic components. Additionally, we propose a novel 1-hot encoding electro-optic analog-to-digital converter (eoADC) architecture to convert the multiplication outputs into digital bitstreams, supporting processing in the electrical domain. Our designed photonic tensor core, utilizing GlobalFoundries' monolithic 45SPCLO technology node, achieves computation speeds of 4.10 tera-operations per second (TOPS) and a power efficiency of 3.02 TOPS/W.
\end{abstract}


\section{Introduction}
The rapid growth of data-intensive applications, such as artificial intelligence (AI), machine learning (ML), and big data analytics, highlights the memory-wall bottleneck in von Neumann systems, where frequent data transfers between the processor and memory reduce speed, bandwidth, and power efficiency \cite{memory_wall_bottleneck}. Addressing this requires a shift in hardware design, exploring new computing paradigms that enhance computational throughput while tackling energy and latency concerns. In-memory computing (IMC), where computation occurs within memory to minimize data transfer delays, is one such approach \cite{imc_ref1, imc_ref2}. Other emerging paradigms, like non-volatile memory (NVM) computing \cite{non_volatile_comp}, optical computing \cite{optical_comp}, and quantum computing \cite{quantum_comp}, can also improve efficiency by bypassing traditional system limitations.

SRAM-based in-memory computing (IMC) macros offer advantages like unlimited endurance and compatibility with existing silicon foundries, but face challenges such as crosstalk between adjacent bitcells during simultaneous row activation, which can lead to read-disturb errors and high power consumption \cite{sram_imc1, sram_imc2, sram_imc3}. Non-volatile memory (NVM) devices like resistive RAM (RRAM) \cite{rram_ref1, rram_ref2}, phase-change memory (PCM) \cite{pcm_ref1, pcm_ref2}, and magneto-resistive RAM (MRAM) \cite{mram_ref1, mram_ref2} offer benefits such as non-volatility and lower power use, but have limitations. PCM and RRAM suffer from slow read/write speeds (nanoseconds to microseconds), high write energy, and limited endurance \cite{rram_ref2, nvm_challenges, nvm_comparison1, nvm_comparison2}, while MRAM, though faster, has a low on-off resistance ratio, making it error-prone and thermally unstable \cite{nvm_challenges, mram_thermal}. Furthermore, both SRAM and NVM approaches depend on electrical interconnects, but as technology scales, these interconnects face significant challenges, including higher coupling capacitance and increased metal wire resistance that exacerbates signal delay and noise issues \cite{interconnect_prb1,interconnect_prb2, comp_imc_sram1, comp_imc_sram2}. These factors collectively limit data throughput and increase energy consumptions for high-speed and energy-efficient computing systems. 

In contrast, photonics systems leverage waveguides to confine and guide light, effectively addressing the scaling challenges faced by electrical interconnects. These systems enable low-loss optical data transfer over significant distances by utilizing carefully designed structures that exploit principles such as total internal reflection, index guidance, or photonic bandgap \cite{optics_advantage1, optics_advantage2}. Moreover, wavelength-division multiplexing (WDM) significantly enhances computational efficiency, bandwidth, and throughput by enabling simultaneous data transmission on multiple wavelengths, each carrying distinct information \cite{multiplex_optical_domain}. Various photonic IMC macros demonstrate ultra-high-speed, energy-efficient operations by encoding inputs as optical pulse intensity and controlling weights via the transmittance or phase of light \cite{photonic_imc_review, photonic_imc, photonic_xbar, photonic_tensor, photonic_imc_pcm, photonic_imc_mzi1, photonic_imc_mzi2, photonic_imc_mzi3, photonic_mrr_weight}. While Mach-Zehnder interferometer (MZI)-based photonic compute cores allow rapid weight updates, their large device area limits scalability for matrix computations \cite{photonic_imc_mzi1, photonic_imc_mzi2, photonic_imc_mzi3}. In contrast, phase-change material (PCM)-based systems offer scalability by controlling transmittance as a weight; however, they demand high write latency and energy, reducing efficiency in large-scale applications requiring frequent updates \cite{photonic_pcm_memory, photonic_imc, photonic_tensor, photonic_imc_pcm}. Microring resonator (MRR)-based photonic tensor cores provide a compact footprint, enabling higher integration density and scalability critical for large-scale matrix compute cores \cite{photonic_xbar, photonic_mrr_weight}. However, MRRs are susceptible to thermal and environmental fluctuations, which can be effectively mitigated through thermal tuning using integrated heaters to stabilize operating conditions \cite{mrm_ref1, mrm_ref2}. Hence, MRR-based photonic IMC systems can deliver high-speed, energy-efficient, scalable, and compact solutions, advancing photonic computing for large-scale applications requiring frequent weight updates.

This work presents a mixed-signal, multi-bit, scalable differential photonic SRAM-embedded tensor core enabling high-speed, energy-efficient matrix multiplication and high-speed memory updates using fabrication-friendly silicon photonics components. A critical challenge in photonic IMC systems—efficient post-processing of computed analog results—is also addressed. Many photonic IMC macros often depend on off-chip processing, such as optical power measurements \cite{photonic_imc, photonic_xbar, photonic_imc_mzi3}, electrical ADCs \cite{photonic_imc_mzi2}, or digital signal processors \cite{photonic_tensor}, which create performance bottlenecks that reduce speed and energy efficiency. To overcome these limitations, we propose a novel monolithic electro-optic ADC (eoADC) architecture. Unlike traditional high-speed flash ADCs \cite{flash_adc1, flash_adc2}, which are power-intensive due to their thermometer-coded design requiring numerous comparator activations, or time-interleaved ADCs, which face synchronization issues and high power consumption \cite{ti_adc1, ti_adc2, ti_adc3}, our eoADC employs a one-hot encoding method. This approach activates only a single thresholding block per conversion, minimizing energy consumption while maintaining flash ADC-level speeds. Additionally, this single-slice design can be extended using a time-interleaved configuration to further enhance speed. By integrating this scalable, energy-efficient eoADC into the photonic tensor core, we provide a seamless, end-to-end architecture compatible with electrical subsystems, advancing large-scale, high-performance computing for big data applications.

The main contributions of this paper are as follows:
\begin{enumerate}
\item We present a novel mixed-signal, multi-bit, and scalable photonic SRAM-augmented photonic tensor core, enabling high-speed, energy-efficient matrix multiplication for complex computational tasks.
\item The architecture supports memory updates at 20 GHz rate, ensuring high-speed operations for big data applications where datasets exceed memory array capacity and require frequent, rapid updates.
\item Additionally, we propose a novel 1-hot encoding electro-optic analog-to-digital converter (eoADC) architecture that utilizes fabrication-friendly silicon photonics components, achieving a speed of 8 GS/s with an energy consumption of 2.32 pJ per conversion.
\item Finally, our photonic SRAM-embedded, multi-bit, mixed-signal photonic tensor core with eoADC demonstrates a computation speed of 4.10 tera-operations per second (TOPS) and a power efficiency of 3.02 TOPS/W.
\end{enumerate}

\section{Photonic Compute Preliminaries}

\begin{figure}[!b]
\centering
\vspace{-0.2 in}
\includegraphics[width=0.6\linewidth]{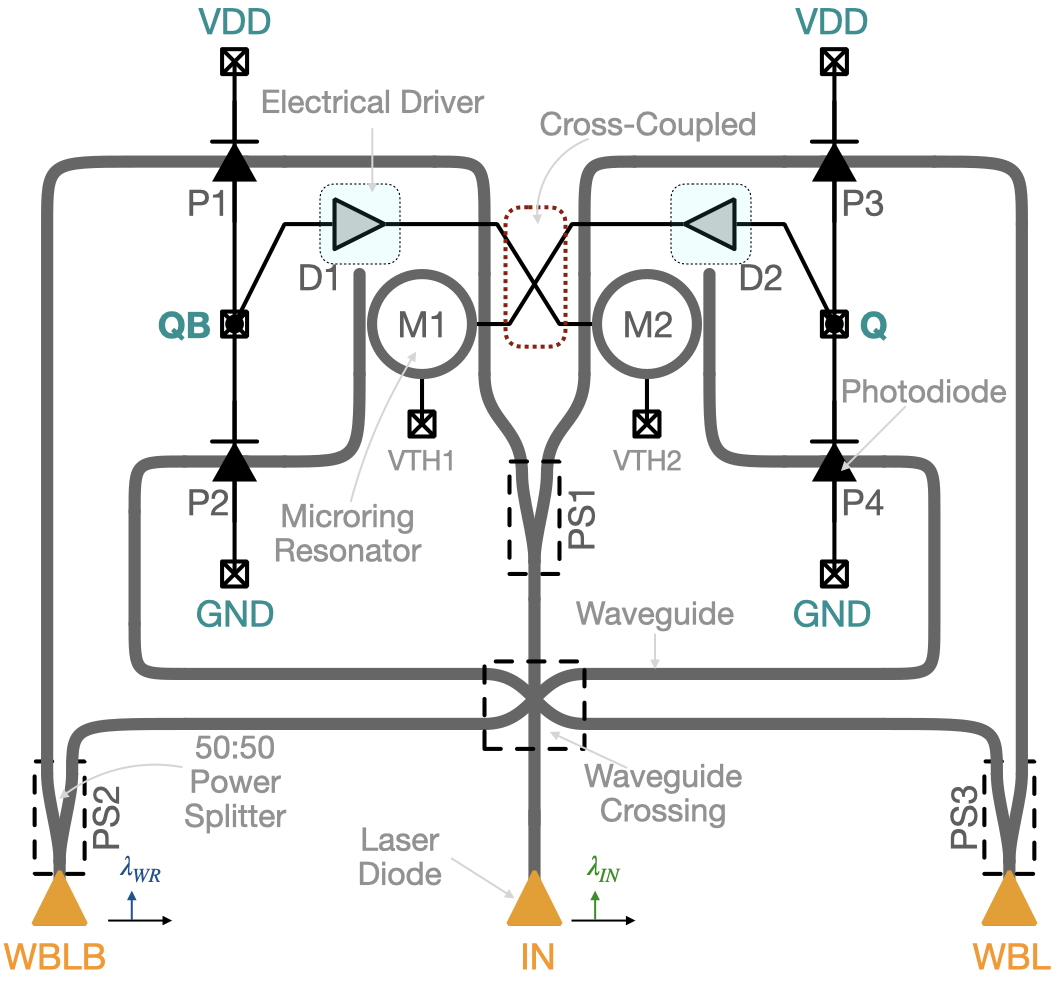}
\caption{Differential cross-coupled photonic SRAM bitcell.}
\label{fig_psram}
\end{figure}

This section explores the core building blocks of the photonic tensor core, leveraging fabrication-friendly silicon photonics components such as waveguides, microring resonators (MRRs), photodiodes (PDs), optical power splitters (PS), and passive absorbers (A). Waveguides confine and guide light through high-refractive-index materials like silicon, with low-refractive-index cladding like silicon dioxide ensuring internal reflection. MRRs consist of a circular waveguide (ring) coupled to one or more straight bus waveguides \cite{mrm_ref1}. When the resonance condition is met—i.e., when the optical path length of the ring is an integer multiple of the input optical wavelength—light couples into the ring; otherwise, it continues through the bus \cite{mrm_ref2}. This resonance depends on factors such as the wavelength of the input light, the effective refractive index of the waveguide mode, and the ring length. Precise tuning of the resonance wavelength can be achieved by modulating the refractive index via an electrical bias across an integrated pn junction (utilizing the plasma dispersion effect). Photodiodes convert optical signals into electrical currents, power splitters divide light among multiple waveguides, and optical absorbers capture stray light to prevent reflections or crosstalk. These mature silicon photonics components utilized to design the photonic SRAM-augmented tensor core and electro-optic ADC, ensuring compatibility with existing silicon processes and integration into electronic platforms. Subsequent sections provide further detail on these photonic blocks.

\subsection{Cross-coupled Differential Photonic SRAM Bitcell:} \label{sec_psram}

Fig. \ref{fig_psram} presents the schematic of the photonic SRAM (pSRAM) bitcell. In this configuration, M1-M2 are micro-ring resonators (MRRs), P1-P4 are photodiodes (PDs), PS1-PS3 are optical power splitters, and A1-A2 are passive optical absorbers to minimize unwanted reflections. An optical laser (\si{\lambda_{IN}}) is connected to the input power splitter (PS1), which directs the power to the input bus waveguides of two identical MRRs (M1 and M2). The wavelength \si{\lambda_{IN}} is selected to resonate with the MRRs when a voltage VDD is applied to them. The thru and drop bus waveguides of M1 (M2) are connected to the waveguides of photodiodes P1 (P3) and P2 (P4), respectively. The midpoints between photodiodes P1 and P2 (P3 and P4) are labeled as QB (Q), serving as the electrical storage nodes of the pSRAM. Node QB (Q) drives M2 (M1) through an  electrical driver D1 (D2), creating a cross-coupled structure to hold the stored data. This arrangement forms a pSRAM latch capable of storing binary data at the storage nodes Q (data) and QB (complementary data) as long as both the optical bias (optical bias via IN) and electrical bias (VDD) are maintained. To retain data, the cross-coupled electro-optic structure must maintain stability. For instance, when Q = 1 (VDD) and QB = 0 (GND) are stored, these values must be preserved as long as the optical and electrical biases are applied. With Q = 1, M1 is tuned to resonance with \si{\lambda_{IN}}, coupling most of the light to P2 via its drop port. This generates a higher photocurrent in P2, creating a low-resistance path to GND and keeping QB at 0. Conversely, QB = 0 shifts the resonance wavelength of M2 away from \si{\lambda_{IN}}. This allows light to pass through M2's thru port to P3, maintaining a high photocurrent and keeping Q at VDD. The complementary states of M1 and M2, coupled with their respective photodiodes, create a positive feedback loop that maintains the stored data. The same mechanism applies in reverse to maintain Q = 0 and QB = 1, where M2 would be on-resonance and M1 off-resonance.

Data can be written into the pSRAM cell by applying differential optical power through the write bitline waveguides (WBL and WBLB). Starting with Q = 0 (GND) and QB = 1 (VDD), to switch the data to Q = 1 and QB = 0, higher optical power is supplied to the WBL waveguide while no power is given to the WBLB. The write optical power must exceed the input bias laser power for successful data flipping and can operate at a different/same wavelength, as photodiodes generally have a broadband response. This causes P3 (P2) to generate more current than P4 (P1), creating a low-resistance path to VDD (GND) for Q (QB), making Q rise to VDD and bringing M1 into resonance, and, QB drops to GND, stabilizing the state. The opposite state (Q = 0, QB = 1) can be written by reversing the optical power between the WBL and WBLB waveguides. More details about the structure and operation of the pSRAM bitcell can be found in \cite{psram_paper}.

\subsection{Mixed-signal Multi-bit Photonic Vector Multiplication Compute Core:} \label{sec_ptc}

\begin{figure}[!t]
\centering
\includegraphics[width=1\linewidth]{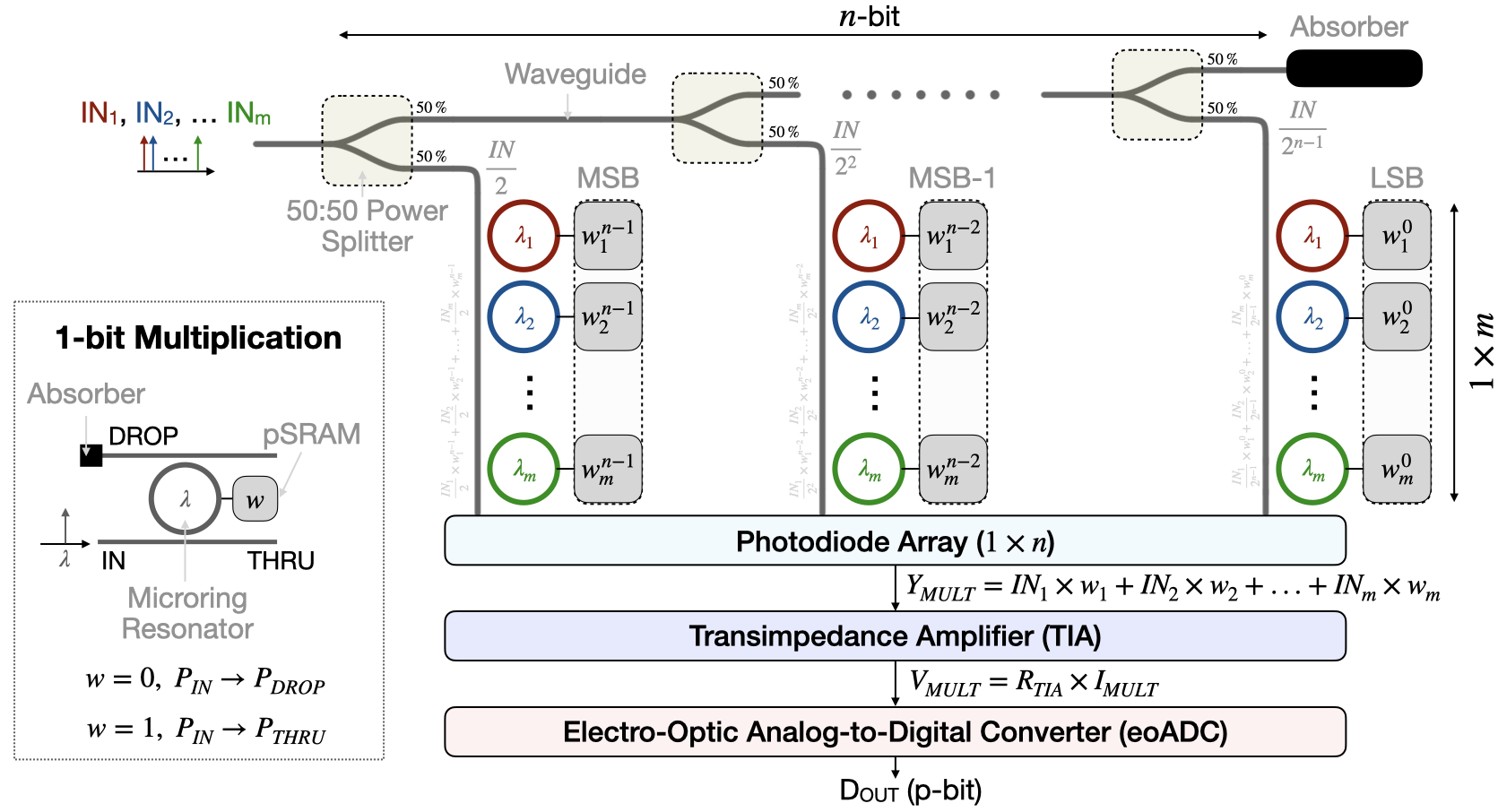}
\caption{Mixed-Signal multi-bit photonic vector multiplication compute core.}
\vspace{-0.2 in}
\label{fig_ptc}
\end{figure}

Fig. \ref{fig_ptc} depicts the mixed-signal, multi-bit (n-bit) vector multiplication compute core, designed for vector-vector multiplication (\si{IN = [IN_1, IN_2, \dots , IN_m]} \si{\times} \si{W = [w_1, w_2, \dots , w_m]}), where \si{IN}, \si{w}, and \si{m} represent the input vector, weight vector, and vector dimension, respectively. The core uses intensity-encoded optical pulses as analog inputs and multi-bit (n-bit) pSRAM (detailed in Section \ref{sec_psram}) for weight storage. In the 1-bit mixed-signal multiplication example in Fig. \ref{fig_ptc}, each unit includes a MRR controlled by a pSRAM bitcell storage node representing a 1-bit weight (w). The MRR's wavelength is tuned such that, when \si{w = 0}, the incoming light is coupled inside the ring, resulting in no output at the thru port (output = 0). Conversely, for \si{w = 1}, the MRR is off-resonance, allowing the light to pass through the thru port. This mechanism enables the multiplication of the analog input (\si{IN}) by the binary weight (\si{w}), producing an output of 0 or \si{IN} based on the weight value.

The analog intensity-encoded vector can be generated using an optical frequency comb, which produces multiple precisely spaced wavelengths \cite{photonic_tensor}, enabling parallel data transmission through Wavelength Division Multiplexing (WDM). The input vector (\si{IN = IN_1, IN_2, \dots , IN_m}) of size \si{m} is transmitted through a single bus waveguide via WDM, with each input intensity encoded at a different wavelength (\si{\lambda_1}, \si{\lambda_2}, \dots , \si{\lambda_m}). For \si{n}-bit MAC operations, \si{n} pSRAMs are assigned to each weight, with weights organized by bit significance (\si{w_1 = w_1^{n-1} w_1^{n-2} \dots w_1^0}), where \si{w_1^{n-1}} is the most significant bit (MSB) and \si{w_1^0} is the least significant bit (LSB). The analog inputs are distributed through \si{n} cascaded power splitters, generating binary-scaled input values (\si{\frac{IN}{2}, \frac{IN}{2^2}, \dots , \frac{IN}{2^{n-1}}})  \cite{psram_sprint}.

The binary-scaled analog signals are multiplied using a 1-bit multiplication structure, which consists of a MRR driven by a 1-bit pSRAM. Each ring is tuned to a specific input wavelength by modulating its length, so each performs multiplication on a corresponding input. For instance, the red, blue, and green rings in Fig. \ref{fig_ptc} perform 1-bit mixed-signal multiplication on \si{IN_1}, \si{IN_2}, and \si{IN_m}, respectively. WDM enables computing across multiple wavelengths within a single bus waveguide without crosstalk. The mixed-signal, wavelength-multiplexed multiplication results are combined within the waveguide and directed to a photodiode. The photodiode array, with \si{n} differential waveguides for n-bit multiplication, aggregates these signals to produce an output equal to the vector-vector multiplication of the analog inputs and n-bit weights. The results are then converted into digital bitstreams (p-bit) using a high-bandwidth transimpedance amplifier (TIA) and ADC.

\subsection{1-hot Encoding Electro-Optic ADC:} \label{sec_padc}

\begin{figure}[!t]
\centering
\includegraphics[width=1\linewidth]{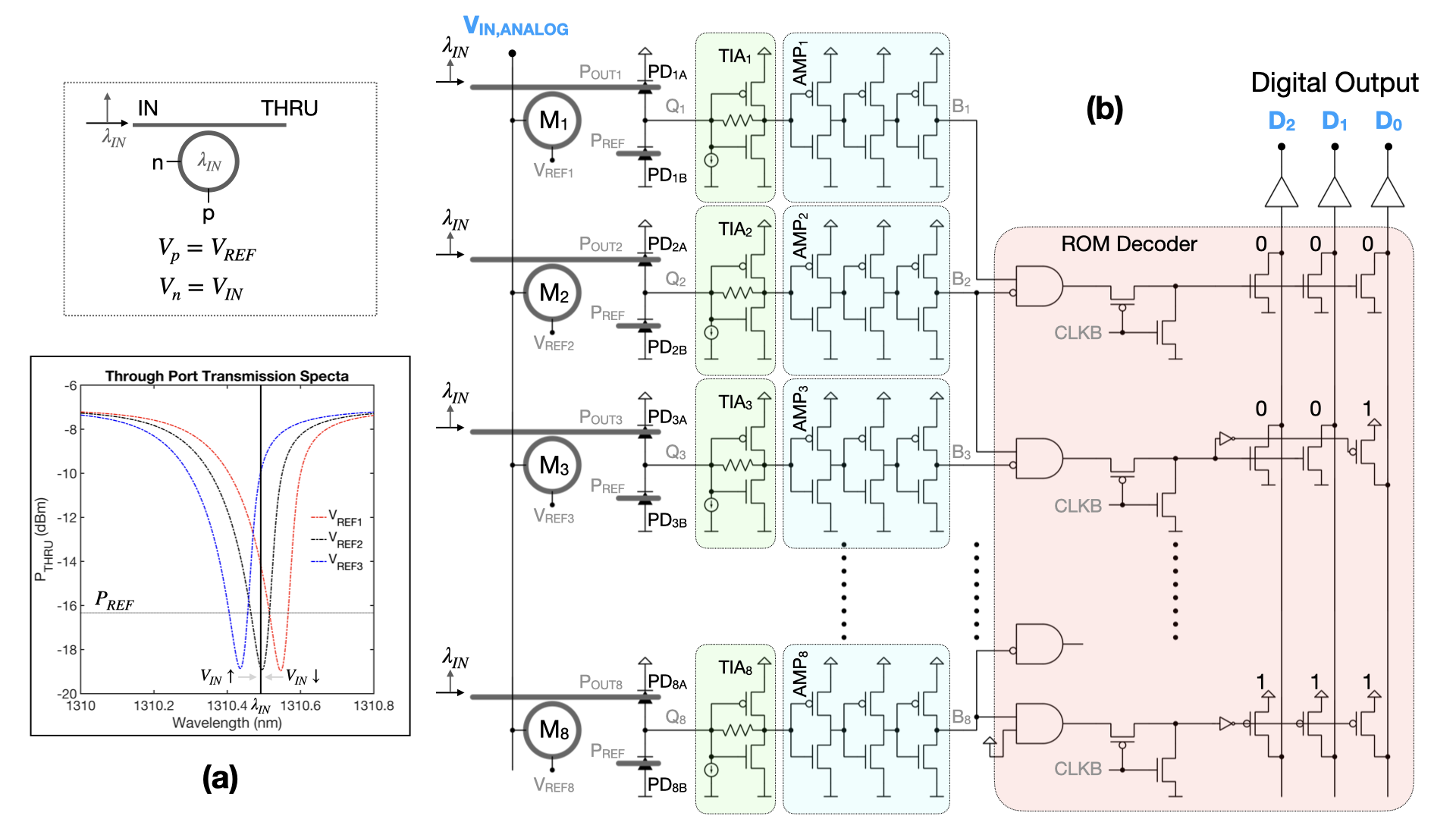}
\caption{(a) MRR transmission spectra as a function of the pn junction voltage, (b) 1-hot encoding electro-optic ADC architecture.}
\vspace{-0.2 in}
\label{fig_padc}
\end{figure}

Fig. \ref{fig_padc}(a) shows the transmission spectra of a two-port MRR, where the resonance state determines whether the thru port receives light. The transmission characteristics are modulated by applying voltage across the pn junction, which controls the refractive index. Three transmission spectra are shown, color-coded as red, black, and blue, corresponding to different applied voltages (\si{V_{REF1} > V_{REF2} > V_{REF3}}) at the p-terminal. In this simulation, \si{V_{IN}} is set to \si{V_{REF2}}, with the resonance wavelength (\si{\lambda_{IN}}) selected when \si{V_{pn} = 0}. At this condition, the MRR's thru port exhibits the lowest power at \si{\lambda_{IN}}, shown by the black curve. When the applied voltage is \si{V_{REF1}} or \si{V_{REF3}}, the MRR remains off-resonance due to \si{V_{pn} > 0} or \si{V_{pn} < 0}, resulting in higher output power at \si{\lambda_{IN}} (\si{> P_{REF}}). As \si{V_{IN}} increases beyond \si{V_{REF2}}, the spectra shift to longer wavelengths (red-shift), and when \si{V_{IN}} reaches \si{V_{REF1}}, the blue curve aligns with the black curve, showing lower output power at \si{\lambda_{IN}}. Similarly, decreasing \si{V_{IN}} shifts the spectra to shorter wavelengths (blue-shift), with the blue curve aligning with the black curve, exhibiting minimum power. This demonstrates that by applying specific reference voltages at the p-terminal and connecting the input to the n-terminal of the pn junction, the MRR can selectively resonate at the input wavelength when \si{V_{IN}} is close to the reference voltage, allowing for ADC quantization at a particular code value.

Leveraging the voltage-dependent notch-like response of the MRR thru port, we present a novel 1-hot encoding electro-optic ADC (eoADC) architecture, shown in Fig. \ref{fig_padc}(b). While the figure illustrates a 3-bit eoADC, the design is scalable to p-bit ADCs. This ADC converts an analog input voltage into 3-bit digital electrical bitstreams, using light as the state variable for high-speed thresholding. For a 3-bit ADC, 8 (for p-bit, \si{2^p}) MRRs are used, with the n-terminals of the pn junctions connected to the input voltage and the p-terminals connected to different reference voltages. This ensures each MRR resonates within a specific input voltage range. For example, MRR \si{M_1} exhibits minimal power at the thru port when the input voltage \si{V_{IN,ANALOG}} is within 0 to \si{\frac{V_{FS}}{8}}, where \si{V_{FS}} is the full-scale ADC input range. Other reference voltages are similarly set to align each MRR to resonate at a distinct input voltage range. By exploiting the transmission spectra and resonance wavelength modulation—governed by the applied voltage across the pn junction—the eoADC achieves 1-hot encoding behavior, distinguishing it from the thermometer-coded approach typically seen in power-hungry flash ADCs. 

A balanced photodiode structure is used as the opto-electric thresholding block, where the lower photodiode connects to a reference optical power (\si{P_{REF}}) and the upper photodiodes are linked to the thru ports of the MRRs, each calibrated to resonate at specific input voltage ranges. When an MRR is on-resonance due to the input voltage, the corresponding upper photodiode receives less optical power than the lower photodiode, causing the output node (\si{Q_{p}}) to discharge toward ground. The output voltage at \si{Q_{p}} is amplified through an inverter-based high-speed TIA and a cascaded voltage amplifier, converting the voltage change into a rail-to-rail swing \cite{optical_rx}. This amplified signal (\si{B_p}) is sent to a ROM-based decoder circuit, which implements a ceiling function between adjacent channels for fast digital bitstream conversion. The ceiling function ensures robustness by resolving cases where the input voltage is at the midpoint of two voltage ranges, preventing simultaneous activation of two digital codes. This avoids static current flow through the decoder, enhancing reliability. The current demonstration utilizes a 3-bit ADC, leveraging the MRR in the GF45SPCLO node; however, higher precision can be achieved by optimizing devices, such as using high-Q MRRs, or by cascading multiple lower-bit ADCs with shift-and-add operations. Additionally, this ADC architecture can also be integrated utilizing time-interleaved structures to improve the operating speed.

\section{Mixed-Signal Multi-bit Scalable Photonic Tensor Core}

\begin{figure}[!b]
\centering
\vspace{-0.1 in}
\includegraphics[width=1\linewidth]{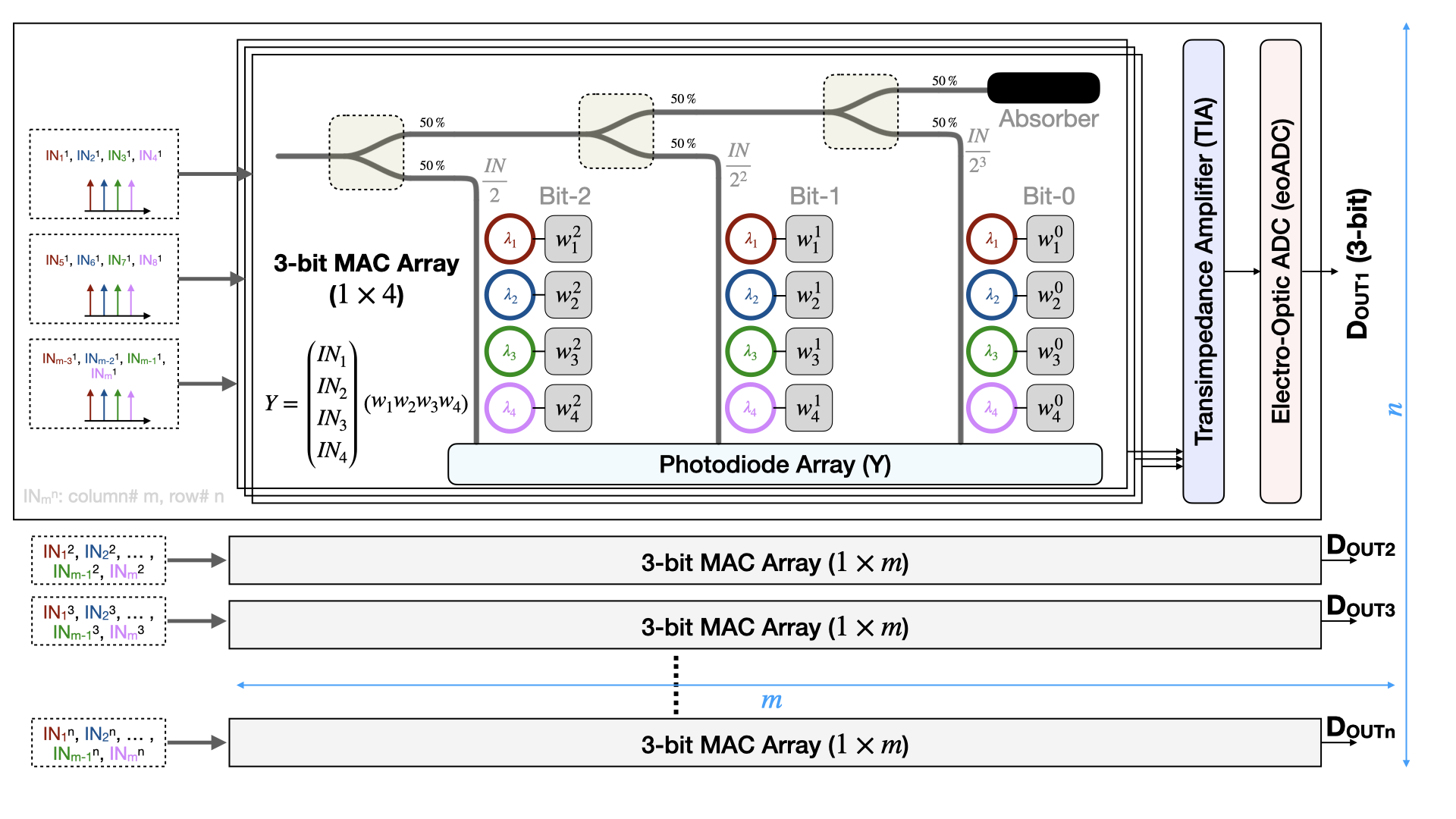}
\vspace{-0.2 in}
\caption{Mixed-signal multi-bit scalable 2D photonic tensor core enabling matrix multiplication.}
\label{fig_ptc2D}
\end{figure}

Fig. \ref{fig_ptc2D} illustrates a scalable 2D mixed-signal, multi-bit photonic tensor core architecture designed for matrix multiplication, which is achieved by tiling the vector multiplication compute core discussed in Section \ref{sec_ptc}. This core employs WDM for mixed-signal multiplication, requiring precise wavelength selection and careful tuning of the MRRs. A key design factor is the number of usable wavelength channels within the MRR’s free spectral range (FSR). For instance, with a 9 nm FSR and 2 nm channel spacing, up to four wavelength channels can be effectively used without causing side-channel interference. Channel spacing can further be lowered to support more wavelength channels depending on the MRR transmission characteristics. In this work, four wavelengths (\si{\lambda_1}, \si{\lambda_2}, \si{\lambda_3}, \si{\lambda_4}) are assigned per vector compute macro, allowing for the multiplication of \si{1 \times 4} input and weight vectors. Specifically, the input elements \si{IN_1}, \si{IN_2}, \si{IN_3}, and \si{IN_4} are multiplied by MRRs tuned to the wavelengths \si{\lambda_1}, \si{\lambda_2}, \si{\lambda_3}, and \si{\lambda_4}, respectively, with control provided by the pSRAM bitcell arrays storing the corresponding weights \si{w_1}, \si{w_2}, \si{w_3}, and \si{w_4}. Although Fig. \ref{fig_ptc2D} shows a 3-bit weight precision, the precision can be enhanced by adding more MRRs and pSRAM bitcells. While the number of wavelengths per compute core can limit the vector size, the architecture can be scaled by replicating the vector compute macro to handle larger vectors (\si{1 \times m}). For example, to perform \si{1 \times 16} vector multiplications, four \si{1 \times 4} vector compute macros can be used, with results obtained through current summation in the photodiodes. These results are then digitized using the eoADC, as explained in Section \ref{sec_padc}. Replicating the \si{1 \times m} compute core \si{n} times allows multiplication for an \si{m \times n} array.

\section{Simulation Results and Performance Analysis}

This section presents the verification results and performance metrics of the building blocks in our proposed photonic tensor core, simulated using the monolithic GF45SPCLO technology node.

\subsection{Weight Configuration in pSRAM Bitcell}

\begin{figure}[!t]
\centering
\includegraphics[width=0.95\linewidth]{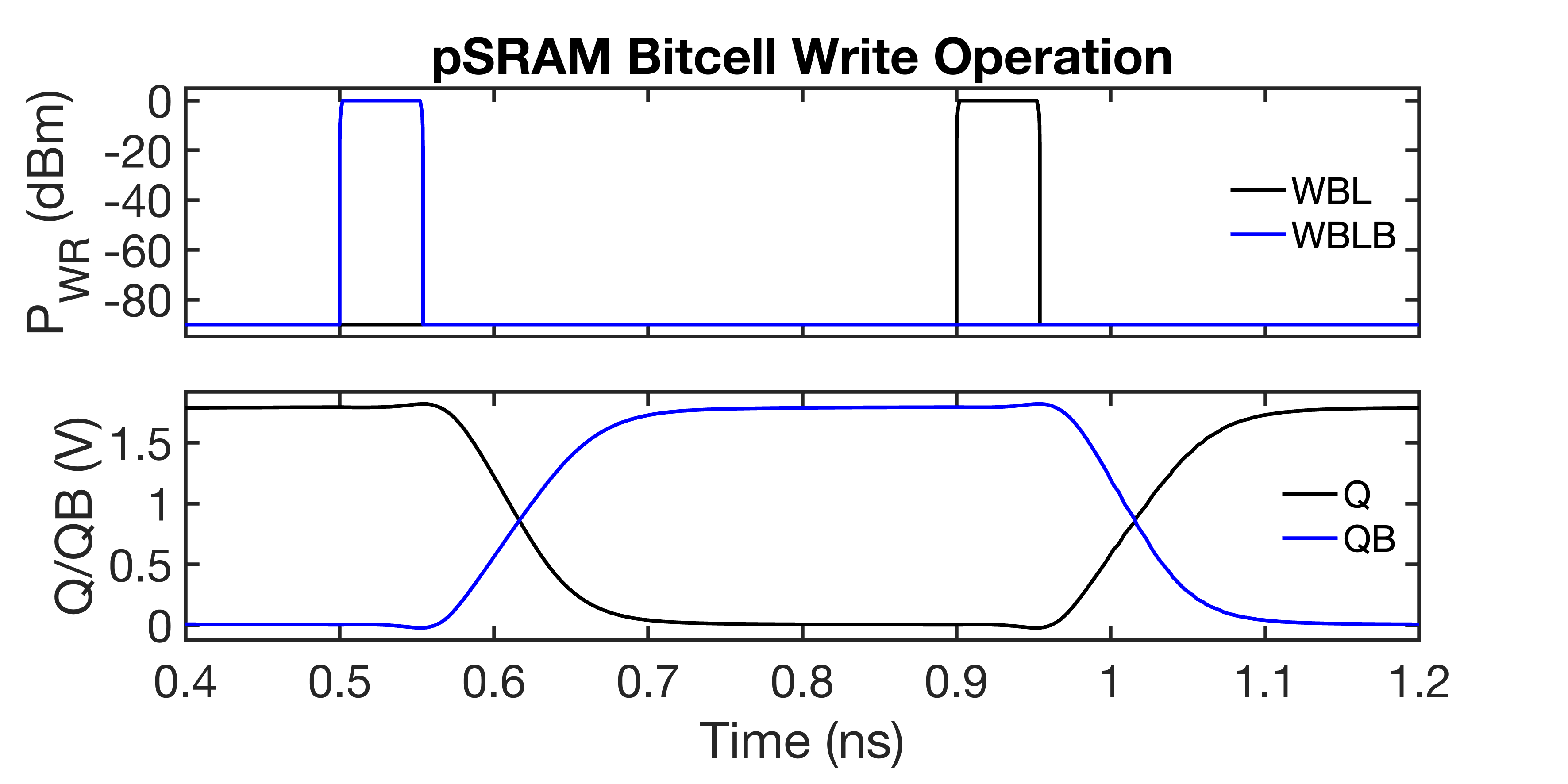}
\caption{Verification of weight configuration in pSRAM bitcell.}
\vspace{-0.2 in}
\label{fig_ver_psram}
\end{figure}

\begin{figure}[!b]
\centering
\vspace{-0.2 in}
\includegraphics[width=0.95\linewidth]{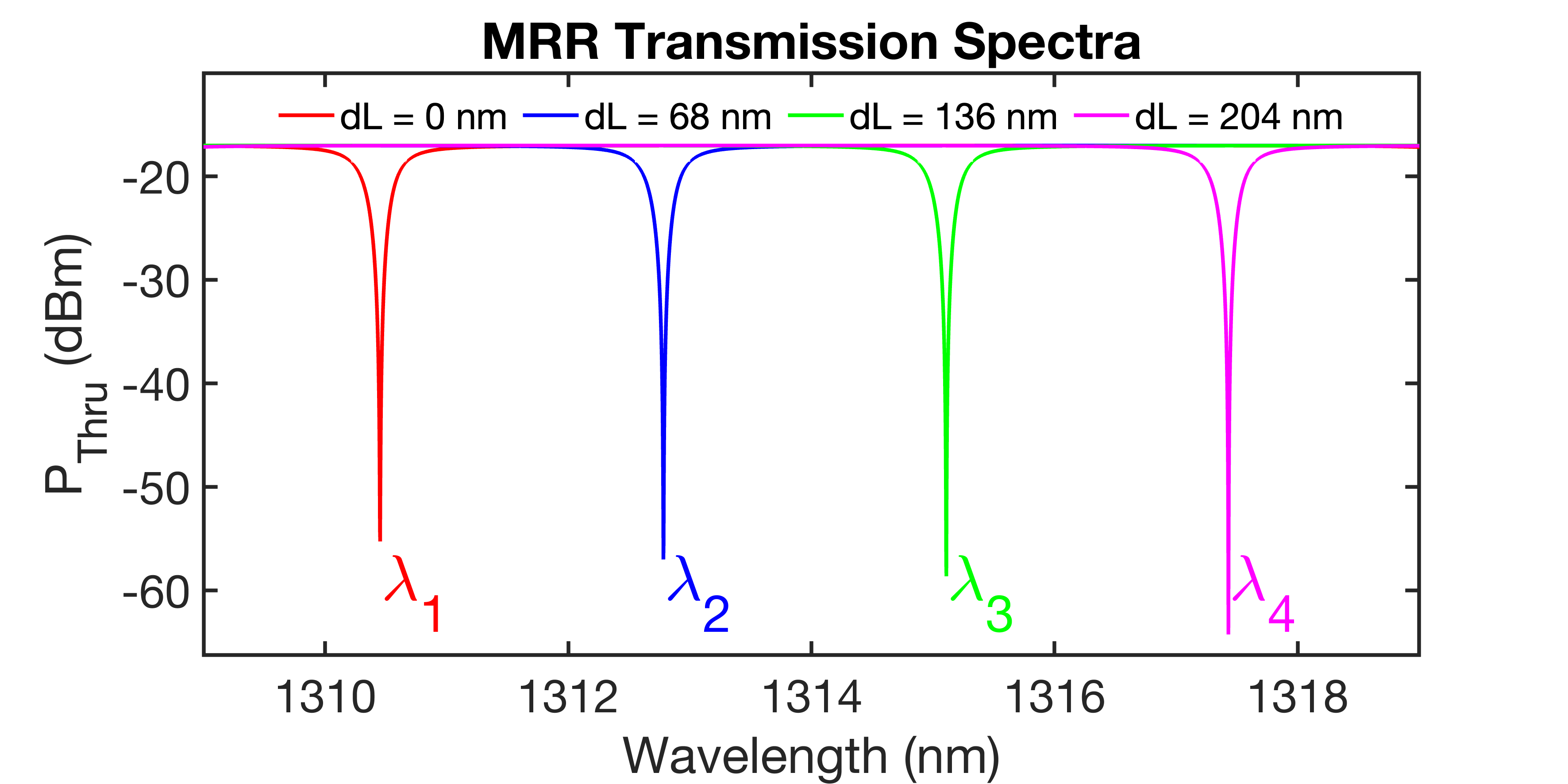}
\caption{Transmission spectra of the MRR as a function the ring adjustment length. Here, dL denotes the adjusted length from the base ring structure.}
\label{fig_ver_wdm}
\end{figure}

To configure weight values, data 0 or 1 is written into the pSRAM bitcell by applying a differential optical pulse via the write bitlines. In Fig. \ref{fig_ver_psram}, the top subplot shows the optical write laser input to the WBL and WBLB waveguides, using a 50 ps wide write pulse at 0 dBm power. The bottom subplot illustrates how storage nodes Q and QB respond to optical inputs at WBL and WBLB, respectively; a write pulse on WBL (or WBLB) sets Q (or QB) to 1. The stabilized (hold mode) Q and QB states after data flip are also shown. With a -20 dBm optical bias and a wall-plug efficiency of 0.23 \cite{wpe} for the input and write laser source, the pSRAM consumes 0.5 pJ of energy per switching event (weight update) at a speed of 20 GHz.

\subsection{Vector Multiplication Compute Core}

\begin{figure}[!t]
\centering
\includegraphics[width=0.95\linewidth]{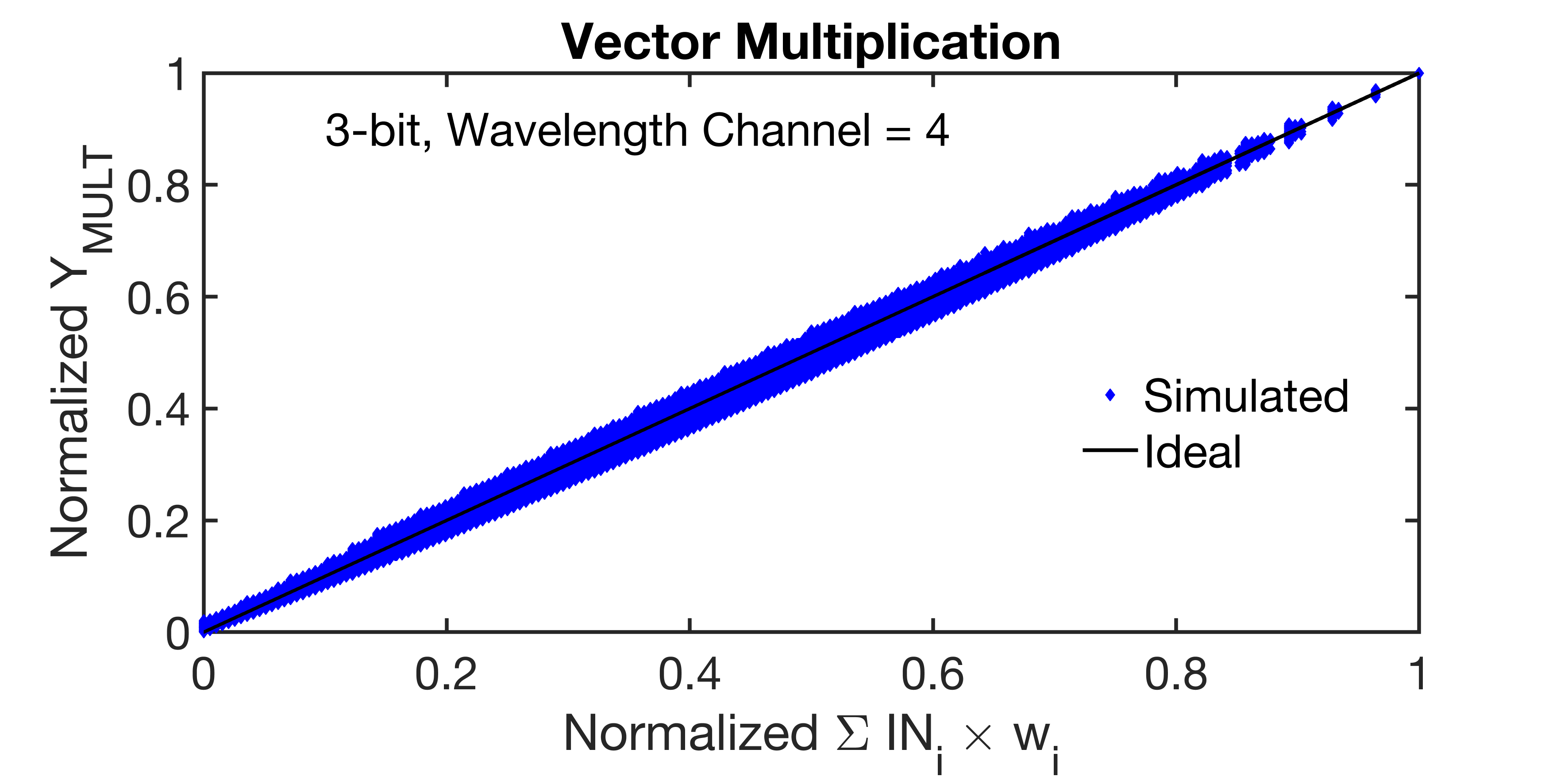}
\caption{Simulation results of multiplying two \si{1\times4} vectors using 3-bit weight precision and four wavelength channels.}
\vspace{-0.2 in}
\label{fig_ver_vector_mult}
\end{figure}

\begin{figure}[!b]
\centering
\vspace{-0.2 in}
\includegraphics[width=0.95\linewidth]{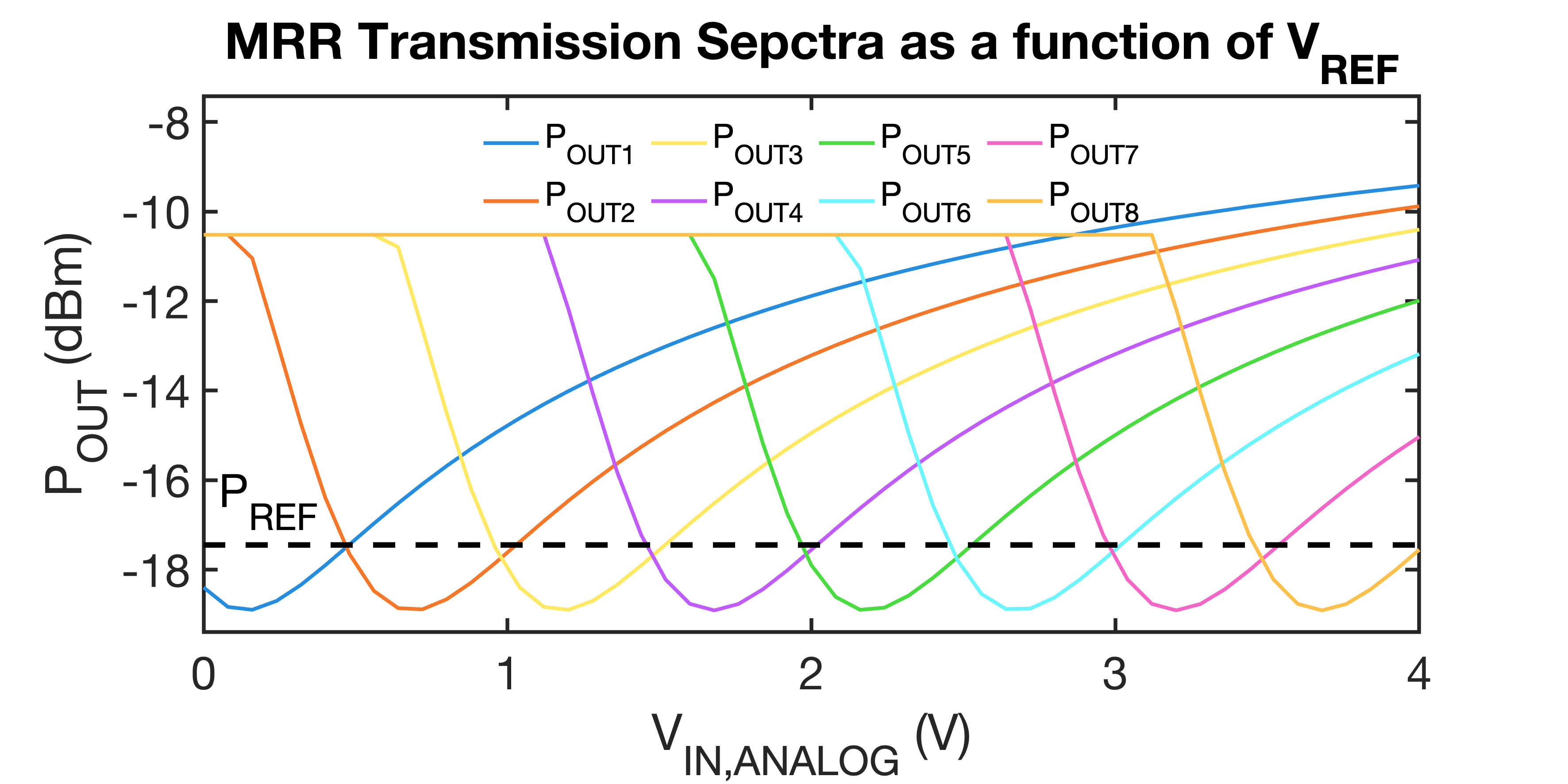}
\caption{MRR transmission spectra versus input analog voltage as a function of different reference voltages (\si{V_{REF}}) exhibiting 1-hot encoding characteristics.}
\label{fig_ver_padc_char}
\end{figure}

Achieving WDM-based multiplication within a single waveguide (as discussed in Section \ref{sec_ptc}) requires precise selection of input and resonance wavelengths. The MRR, with a 7.5 \si{\micro\meter} ring radius and a 200 \si{\nano\meter} gap at the thru-port, achieves four distinct resonance wavelengths (\si{\lambda_1}, \si{\lambda_2}, \si{\lambda_3}, and \si{\lambda_4}) by adjusting the ring length by 0 nm, 68 nm, 136 nm, and 204 nm, respectively. With an FSR of 9.36 nm and a wavelength separation of 2.33 nm, minimal crosstalk is ensured. While more wavelengths could fit within the FSR by reducing the separation, this work focuses on using four wavelengths for multiplication on four inputs within the same waveguide.

Fig. \ref{fig_ver_vector_mult} illustrates the simulation results for multiplying two \si{1 \times 4} vectors, where the input values are represented by analog light intensities, and the weights are encoded using 3-bit pSRAM bitcells. The simulation leverages four wavelength channels, though the GF45SPCLO node supports simulation for only one wavelength at a time. To address this, each wavelength channel is simulated separately, with all the MRRs included in the testbench to incorporate the inter-channel crosstalk, and the results are combined linearly through photodiode current summation to generate the final output. Ideally, the normalized photodiode current output should align linearly with the vector multiplication results, and the simulated outputs follow this trend, as shown in the figure.

\subsection{Electro-optic ADC Verification}

Fig. \ref{fig_ver_padc_char} shows the MRR thru-port transmission spectra (\si{P_{OUT}}, as in Fig. \ref{fig_padc}) of the electro-optic ADC. Each MRR (\si{M_1} to \si{M_8}) corresponds to a distinct reference voltage, creating dips in the transmission spectra within specific input voltage ranges. As the input voltage approaches a reference voltage, the corresponding MRR generates a dip. For a given input code width (1 LSB of the ADC), only one transmission spectrum produces power lower than the reference, activating a single opto-electric thresholding block (1-hot encoding) as a function of the input voltage.

\begin{figure}[!t]
\centering
\includegraphics[width=0.95\linewidth]{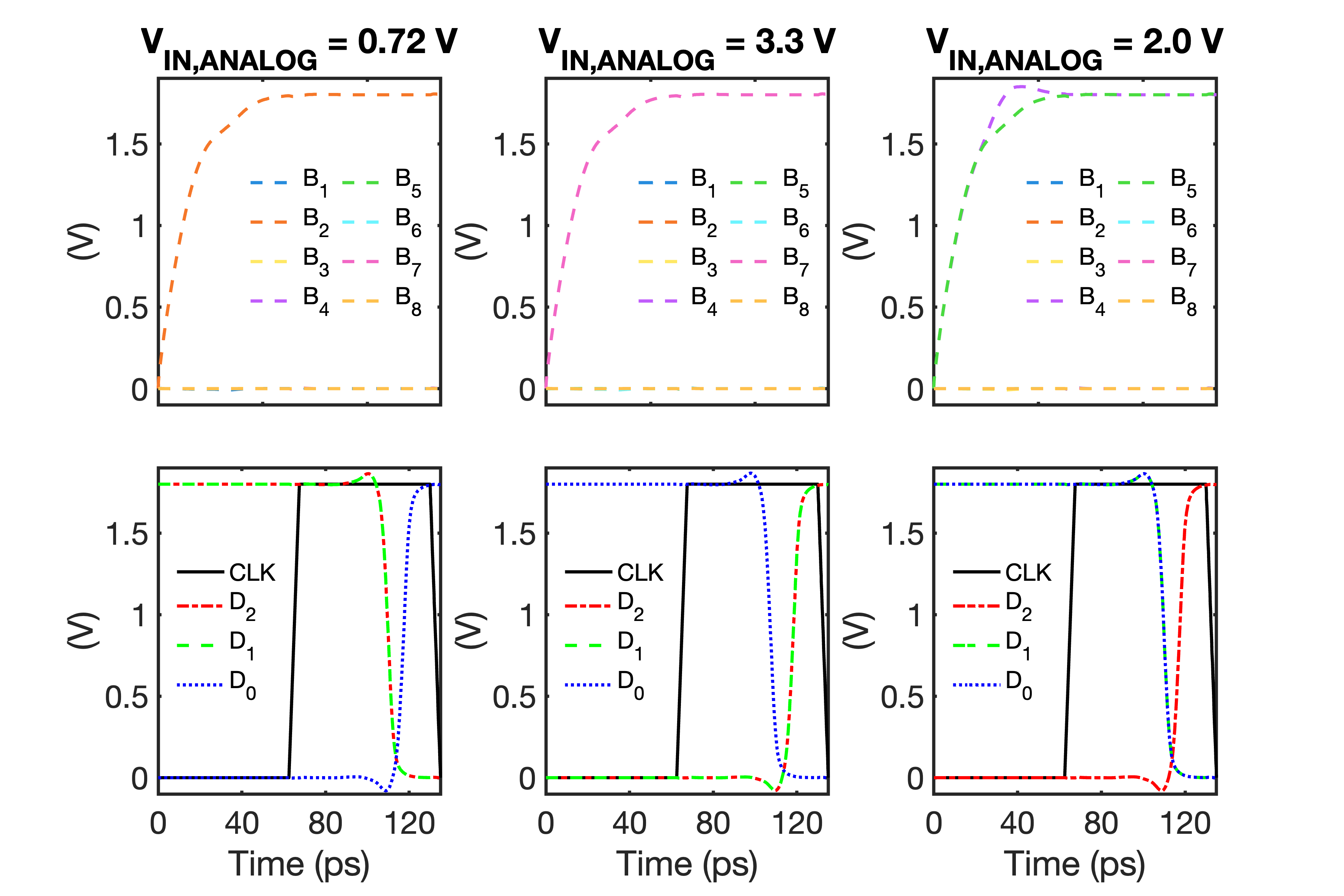}
\caption{Transient verification results of the eoADC architecture.}
\vspace{-0.2 in}
\label{fig_ver_padc_tran}
\end{figure}

\begin{figure}[!b]
\centering
\vspace{-0.2 in}
\includegraphics[width=0.95\linewidth]{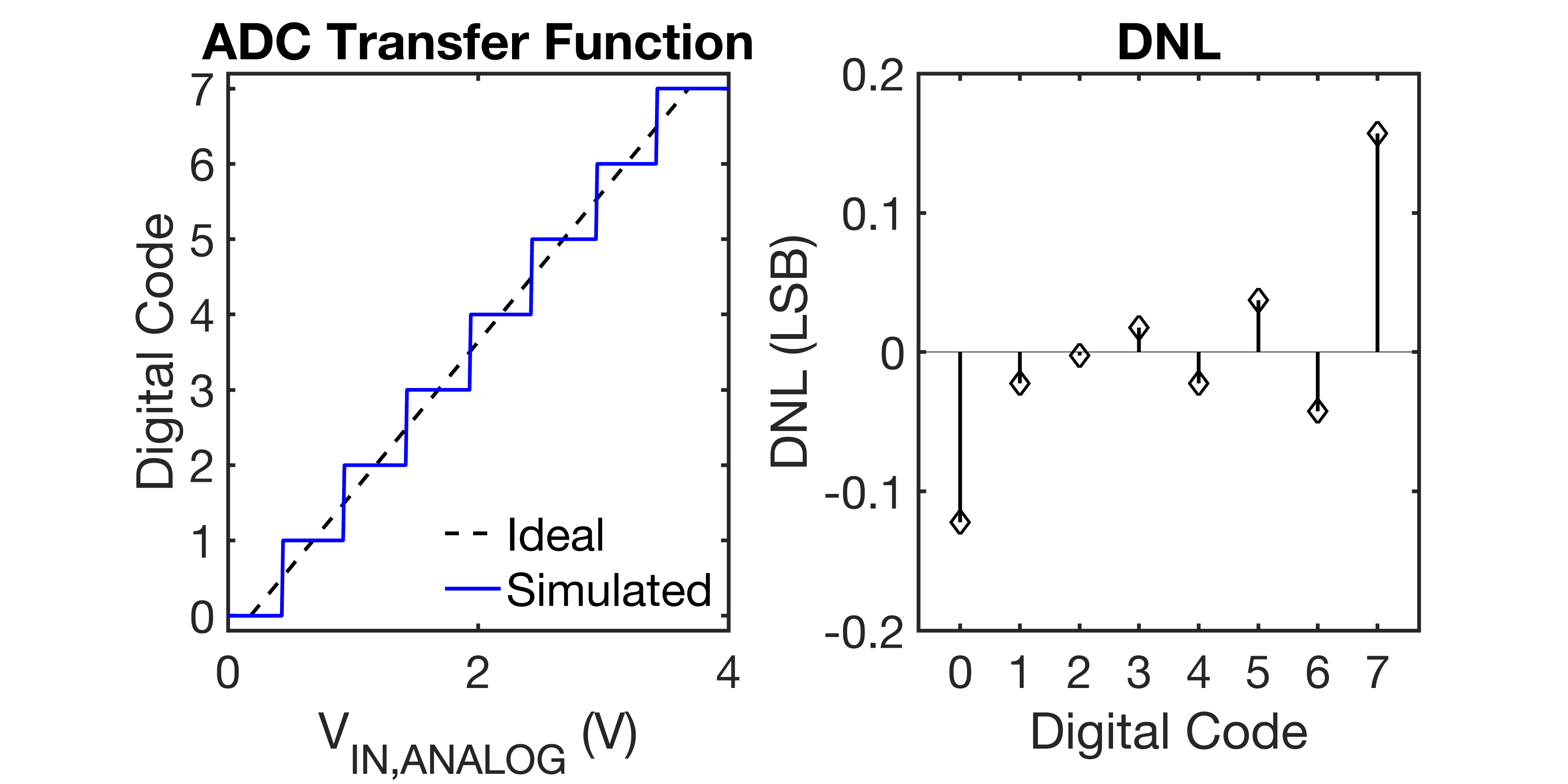}
\caption{ADC transfer function (left-subplot) and differential nonlinearity (DNL) characteristics (right-subplot).}
\label{fig_ver_padc_tf_dnl}
\end{figure}

Fig. \ref{fig_ver_padc_tran} shows the transient characteristics of our proposed eoADC for three input settings. The subplots in the top row indicate that for analog inputs of 0.72 V and 3.3 V, only one opto-electric thresholding block, followed by the TIA and amplifier (\si{B_2} and \si{B_7}, respectively), is activated. The ROM-based decoder outputs the digital codes \si{001} and \si{110}, as seen in the bottom subplots. For an analog input of 2 V, two activations (\si{B_4} and \si{B_5}) occur since 2 V lies at the boundary between two adjacent codes. However, the ceiling priority ROM-based decoder correctly outputs \si{100}, demonstrating the accuracy, robustness, and reliable operation of our eoADC at a sampling speed of 8 GS/s (\si{\sim} 125 ps clock period).

Fig. \ref{fig_ver_padc_tf_dnl} shows the simulation results of the ADC transfer function and differential non-linearity (DNL). The code width closely matches the ideal, with no missing codes (no DNL of -1 LSB). A 10 \si{\micro\meter} radius MRR with a 250 nm gap was used to achieve 1-hot encoding, with 200 \si{\micro W} input optical power at 1310.5 nm and 18 \si{\micro W} optical reference power per channel. The ADC operates with 1.8 V analog and digital supply voltages. With a wall-plug efficiency of 0.23 \cite{wpe}, the total optical power is 7.58 mW, and the total electrical power is 11 mW. Eliminating the cascaded amplifiers and TIAs can reduce power consumption but results in slower speed, e.g., the eoADC without these components operates at 416.7 MS/s while consuming 58\% less electrical power. In addition, optimizing the MRR's voltage modulation efficiency can improve bit precision and speed.


\begin{table}[!t]
\caption{Performance comparison of various Photonic IMC Macros.}
\label{comparison}
\begin{center}
\begin{tabular}{|c|c|c|c|} \hline
Reference & Throughput & Power Efficiency & Weight Update  \\
          & (TOPS)     & (TOPS/W)        &  (Speed)     \\ \hline
\cite{photonic_imc_mzi2}  & 0.12 & ---   & 60 GHz \\
\cite{comp_photonic_imc1} & 0.93 & 0.83  & $<$ 0.5 GHz$^*$ \\
\cite{comp_photonic_imc2} & 11.0 & ---   & 2 Hz$^\dagger$ \\
\cite{comp_photonic_imc3} & ---  & 10    & $\sim$ 1 GHz $^\ddagger$ \\
\cite{comp_photonic_imc4} & 3.98 & 1.97  & $<$ 0.5 GHz$^*$  \\
\textbf{This Work} & 4.10 & 3.02 & 20 GHz\\
\hline
\multicolumn{4}{l}{$^*$ \scriptsize{FPGA-controlled multi-channel DC power supply}} \\
\multicolumn{4}{l}{$^\dagger$ \scriptsize{Utilizing Finisar WaveShaper 4000S, settling time 500 ms}} \\
\multicolumn{4}{l}{$^\ddagger$ \scriptsize{PCM write speed}}
\vspace{-0.3 in}
\end{tabular}
\end{center}
\end{table}

\subsection{Performance Analysis}

A \si{16 \times 16} photonic tensor core was analyzed to evaluate performance metrics, enabling it to compute 16 vector multiplications of two \si{1 \times 16} vectors. With 3-bit weight precision, the core incorporates 768 pSRAM bitcells. Four wavelength channels are used for WDM-based multiplication, allowing 4 vector multiplication cores to simultaneously process \si{1 \times 16} vector multiplications per row. While the MRRs and photodetectors offer high electro-optical bandwidth for fast operations, latency from the electro-optic ADC limits the overall speed. The core achieves a computational throughput of 4.10 TOPS (1 operation = 3-bit multiplication/addition). Power efficiency is calculated considering the pSRAM bitcells, mixed-signal multiplication cores, electro-optic ADC, TIA \cite{tia_power}, and a laser wall-plug efficiency of 0.23 \cite{wpe}, yielding 3.02 TOPS/W. Table \ref{comparison} compares throughput, power efficiency, and weight update speeds across various photonic IMC macros.

\section{Conclusion}

In summary, we present a novel, mixed-signal, multi-bit, scalable photonic SRAM-augmented tensor core that enables ultra-fast, energy-efficient matrix multiplication computations. The proposed architecture supports multi-GHz memory updates, suitable for large-scale datasets and in-situ training. It leverages fabrication-friendly integrated silicon photonics utilizing GlobalFoundries' monolithic 45SPCLO technology node, enabling seamless integration into existing foundries. Additionally, our novel high-speed, energy-efficient 1-hot encoding electro-optic ADC architecture enables an end-to-end photonic tensor core that is seamlessly compatible with electrical subsystems. Consequently, leveraging the speed, energy efficiency, and bandwidth advantages of photonics, this architecture establishes a robust platform for next-generation, high-performance, and data-centric computing systems.

\section*{Acknowledgment}
This work is supported by the Defense Advanced Research Projects Agency (DARPA) under Grant No. HR001123S0024.

\newpage
{\small
\bibliographystyle{ieee_tran}
\bibliography{ref}
}

\end{document}